\documentclass[conference]{IEEEtran}

\AtBeginDocument{%
    \providecommand\BibTeX{{%
        \normalfont B\kern-0.5em{\scshape i\kern-0.25em b}\kern-0.8em\TeX}}}

\usepackage{
    latexsym,
    color,
    xcolor,
    multirow,
    array,
    listings,
    enumitem,
    multicol,
    balance,
    graphicx,
    hyperref,
    url,
    algorithmic,
    textcomp
}
\usepackage{amsmath,amsfonts}
\usepackage{fancyvrb}

\usepackage[linesnumbered,ruled,vlined]{algorithm2e}

\newcolumntype{P}[1]{>{\centering\arraybackslash}p{#1}}
\newtheorem{auxdefn}{Definition}[section]

\newtheorem{auxexample}{Example}[section]

\usepackage{flushend}

\begin{document}

\title{JITScope: Interactive Visualization of JIT Compiler IR Transformations}

\author{
    \IEEEauthorblockN{Kyra Dalbo \hspace{7.9cm} Yumna Ahmed}
    \IEEEauthorblockA{
        \begin{tabular}{c@{\hspace{3cm}}c}
        \textit{Department of Mathematics and Computer Science} & \textit{Department of Mathematics and Computer Science} \\
        \textit{Davidson College} & \textit{Davidson College} \\
        Davidson, USA & Davidson, USA \\
        kydalbo@davidson.edu & yuahmed1@davidson.edu
        \end{tabular}
    }
    \\
    \IEEEauthorblockN{HeuiChan Lim}
    \IEEEauthorblockA{
        \textit{Department of Mathematics and Computer Science} \\
        \textit{Davidson College} \\
        Davidson, USA \\
        telim@davidson.edu
    }
}



\maketitle

\begin{abstract}
The complexity of modern Just-In-Time (JIT) compiler optimization poses significant challenges for developers seeking to understand and debug intermediate representation (IR) behavior. This work introduces JITScope, an interactive visualization framework that illustrates how IR nodes and instructions evolve across compilation phases. The system features a full-stack architecture: a Python-based backend transforms raw JSON-formatted IR data—representing an abstract model of the JIT compiler IR—into a normalized SQLite database; a controller layer serves processed CSV data; and a D3.js-powered frontend renders an interactive, phase-aware graph of IR node transformations. The design emphasizes modularity, traceability, and flexibility. Our roadmap explores intuitive visual representations of phase-level changes in IR node connectivity, values, and access patterns. Ultimately, JITScope lays a foundation for future tooling that enables visual exploration of IR evolution, including phase filtering, value tracking, and function-access mapping—offering a new lens into the behaviors and impacts of compiler optimizations.
\end{abstract}


\IEEEpeerreviewmaketitle

\newcommand{\smallfont}{\fontsize{8pt}{10pt}\selectfont}
\newcolumntype{P}[1]{>{\centering\arraybackslash}p{#1}}

\newcommand{\Entities}[1]{\ensuremath{{\cal E}_{P}(#1)}}

\section{Introduction}\label{sec:introduction}
We propose JITScope, a system that visualizes the evolution of Just-in-Time (JIT) compiler's Intermediate Representation (IR) using a backend-driven, phase-aware, graph-based visualization. This paper outlines this visualization framework's architecture, design decisions, challenges, and future directions.

JIT compilers are a class of compilers that dynamically translate and optimize bytecode into machine instructions at runtime to improve the performance of code execution. JIT compilers are widely used in many modern systems, including web browsers~\cite{meurer2016v8,chakracore,javascriptcore,ionmonkey} to operating system kernels~\cite{greg2019bpf}, virtual machines such as the Java Virtual Machine (JVM)~\cite{oracleJVM2024}, and language runtimes like PyPy~\cite{pypy}.

JIT compilers construct and transform a graph known as IR to optimize the input bytecode. Therefore, understanding the transformations applied to IR during JIT compilation is important for compiler engineers, researchers, and developers debugging or enhancing the optimization. However, the data structures and logic governing these transformations are often deeply nested, non-transparent, and challenging to trace across multiple compiler phases.

To help developers quickly analyze the behavior of JIT compilers during optimization, several research efforts have been made. In particular, Lim and Debray~\cite{lim2022modeling,lim2021buglocalization,lim2023dynamicbugs} proposed an approach that focuses specifically on analyzing IR-level optimizations. Their work involves constructing dedicated models to localize bugs that arise during these optimization phases. However, their approach relies on the generated models solely to identify the relevant, i.e., those most likely to contain bugs, function(s) in the JIT compiler’s source code and produces text-based ranking reports. These reports provide limited flexibility for developers to investigate further or interactively explore the underlying optimization behavior. Once the final reports are generated, the models are no longer required or used, making the analysis process largely static and non-exploratory.

Existing approaches for visualizing a system’s execution behavior often focus on complete execution traces~\cite{deelen2006visualization}, call graphs~\cite{intel2007vtune}, control-flow graphs (CFGs)~\cite{Mikhailov2016cfg}, or data flow~\cite{cornelissen2008framework}. JIT compilers, unlike static (ahead-of-time) compilers, are typically integrated into larger runtime systems. For example, JavaScript engines include a JIT compiler as part of a broader architecture that also features a bytecode interpreter and additional components for monitoring and optimizing code execution performance. Thus, while existing approaches that visualize a system’s overall execution can offer high-level insights, they are not well-suited for diagnosing issues that arise specifically within JIT compilers. This is because JIT compilation involves complex, phase-specific transformations of IRs that are not captured in standard visualizations. As a result, developers and researchers lack the fine-grained, phase-aware views needed to trace subtle bugs, understand missed optimizations, or reason about dynamic code generation—making traditional visualizations insufficient for JIT-focused analysis.

A closely related effort in visualizing JIT compiler IRs is the work by Lim and Kobourov~\cite{lim2021jitgraphs}. Their visualization is based on IR models generated using Lim’s earlier research on modeling and localizing JIT compiler bugs~\cite{lim2022modeling,lim2021buglocalization,lim2023dynamicbugs}. The goal of their approach is to present the entire IR structure in a single, coherent view using the metro map metaphor~\cite{jacobsen2021metrosets}, rather than visualizing the IR’s transformation across individual compilation phases. While the layout is visually clean and offers a high-level summary of the IR, the visualization lacks support for showing how the IR evolves over time, limiting its usefulness for phase-by-phase analysis and practical debugging tasks.

Our research aims to provide a more interactive and phase-aware visualization of JIT compilation behavior. With JITScope, we focus on capturing the evolution of the IR across multiple compiler phases rather than presenting a static snapshot. Our approach includes the development of a backend SQLite database that parses the deeply nested JSON format IR graph data, a controller layer that converts SQL queries into JSON outputs, and a D3.js-based front-end graph that will ultimately enable interactive phase selection, node tracking, and instruction-level exploration. While we have not finalized the full front-end deployment, our current prototype reflects the functional goals and design thinking behind JITScope. This paper presents the architecture, design decisions, and key challenges in building JITScope, and outlines directions for extending its capabilities.

The remainder of this paper is organized as follows: (1) We begin by introducing key background concepts to help readers better understand our approach and the broader context of JIT compilers; (2) We then present the design and architecture of our system; (3) Next, we outline our evaluation plan, which we will carry out once the system is fully implemented; (4) We discuss both the challenges encountered during development and potential issues that may arise in the future; (5) We then review closely related work; and (6) We conclude the paper with a summary and future directions.


\section{Background and Motivation}\label{sec:background}
\subsection{Just-in-Time (JIT) Compilers}
\begin{figure}[h]
\centering
\includegraphics[width=\columnwidth]{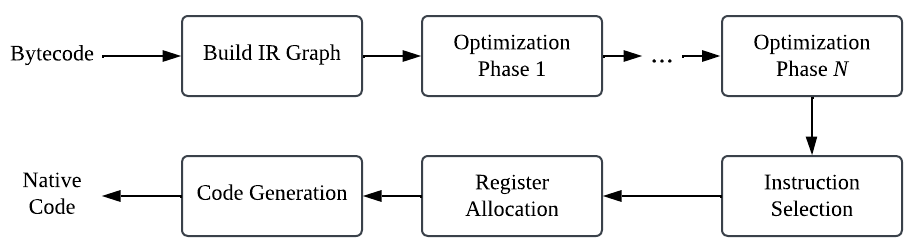}
\caption{Overview of JIT Compiler Architecture}
\label{fig:jitcompiler_pipeline}
\end{figure}
A Just-in-Time (JIT) compiler is a runtime component of a larger system—such as a JavaScript engine or the Java Virtual Machine (JVM)—that dynamically optimizes and translates bytecode into native machine code~\cite{ibmJIT2025}.

Figure~\ref{fig:jitcompiler_pipeline} shows the overview of JIT compiler architecture, which was adapted and modified from Ishizaki \emph{et al.}~\cite{ibmJIT2025} paper on Java JIT compiler. The JIT compiler receives bytecode from the interpreter or runtime environment and begins by constructing an intermediate representation (IR) in the form of a graph.

Unlike static ahead-of-time (AOT) compilers, where optimization is often optional, JIT compilers are designed to optimize performance-critical code dynamically. They apply a series of optimization passes—such as method inlining, dead code elimination, and constant propagation—which transform the IR across multiple phases. Once the IR has been fully optimized, the back-end of the JIT compiler generates native machine code by selecting architecture-specific instructions and performing register allocation.

Understanding how the IR evolves through the optimization phases is challenging, as internal representations are typically neither visualized nor exposed in a queryable form. In particular, JIT compilers, like AoT compilers, may include a large number of optimization phases—potentially hundreds~\cite{wikipediaOptimizingCompiler}—with each phase responsible for transforming the IR graph in specific ways.

\subsection{Intermediate Representation (IR)}
Intermediate Representation (IR) is a graph, where the nodes represent operations or data and the edges represent the relationship, e.g., dependencies, between the nodes. Different JIT compilers adopt different graph structures for their IRs. For example, Google’s V8~\cite{meurer2016v8} and Java’s HotSpot JVM~\cite{oracleJVM2024} JIT compilers use the sea-of-nodes~\cite{demange2018seaofnodes} representation, Mozilla’s IonMonkey~\cite{ionmonkey} employs a traditional control-flow graph (CFG), and Apple’s JavaScriptCore~\cite{javascriptcore} JIT utilizes a data-flow graph (DFG) structure.

In our work, we adapted Lim and Debray's IR model~\cite{lim2022modeling,lim2021buglocalization,lim2023dynamicbugs}, which is the data that Lim and Koborouv~\cite{lim2021jitgraphs} used to visualize the JIT compiler IRs in the metro metaphor~\cite{jitcompilerirviz}. The IR data is stored in a JSON-formatted file that contains a list of IR nodes and a mapping from JIT function IDs to the corresponding compiler source function symbols that executed transformations on the IR. Each node in the file represents a JIT compiler IR node and includes the following information:

\begin{itemize}
    \item The memory address of the node.
    \item The opcode associated with the node.
    \item A list of opcode updates applied to the node.
    \item The set of edges (i.e., connections to other nodes).
    \item The list of values held by the node.
    \item A boolean flag indicating whether the node is alive at the end of optimization (i.e., $\textit{true}$ if alive, $\textit{false}$ if removed).
    \item A log of all instructions that accessed the node during optimization. Each instruction is tagged with a unique instruction ID and a function ID, which can be mapped back to the corresponding source-level function symbol.
\end{itemize}


\section{Design Plan}\label{sec:research}
\begin{figure}[h]
\centering
\includegraphics[width=0.7\columnwidth]{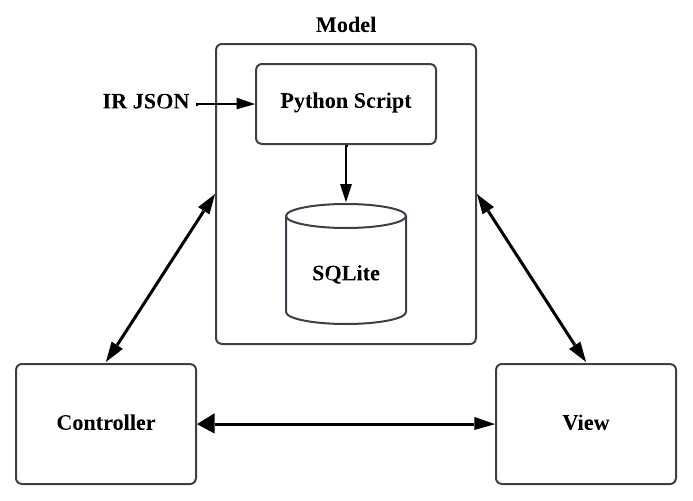}
\caption{Overview of JITScope Architecture}
\label{fig:jitscope_pipeline}
\end{figure}
\subsection{System Pipeline}
Our system adopts the traditional Model-View-Controller (MVC) architecture. Figure~\ref{fig:jitscope_pipeline} illustrates an overview of the system's structure. The input to the system is a JSON-formatted file containing IR data, uploaded by the user. Upon upload, a Python script processes the file by extracting relevant information and loading it into a normalized SQLite database. The controller handles database queries to generate CSV files representing the IR data. It is also responsible for validating user input and passing it to the model. The front-end, built with D3.js, reads these CSV files to render interactive node-link diagrams, enabling users to visually and contextually explore the transformations of the intermediate representation.

While it may seem sufficient to use a standalone Python script to convert the input JSON file to CSV, our use of the MVC architecture offers several important advantages. First, MVC promotes a clean separation of concerns: the model manages the data and logic, the view handles the visual representation, and the controller manages communication between them. This separation makes the system more maintainable, testable, and scalable. Second, by centralizing data handling in the model and controller, our system becomes more adaptable to changes in the view layer. For example, if we later decide to switch from a D3-based front-end to another visualization framework, we can do so without modifying the data extraction logic. Similarly, enhancements to the model—such as supporting additional IR formats or new filtering options—can be made independently of the front-end. In contrast, a monolithic Python script hardwires data transformation to a specific output format, making future changes or extensions more difficult to manage.


\subsection{JSON-to-Database}
The conversion from JSON to SQLite is performed by a dedicated Python script that systematically parses the hierarchical structure of the input data. This script extracts node-level metadata, constructs edge relationships, and populates auxiliary information such as function identifiers, etc. A notable challenge addressed in this process is the assignment of phase names to instruction-node access records. This is achieved by traversing function ID ranges and mapping them to their corresponding optimization phases. All transformations—whether opcode changes, value updates, or edge modifications—are normalized into dedicated relational tables, with records consistently linked via foreign key constraints. Once the data is fully cleaned, structured, and validated, it is committed to the SQLite database, making it ready for downstream querying and visualization.

\subsection{Database Schema}
\begin{figure}[h]
\centering
\includegraphics[width=0.7\columnwidth]{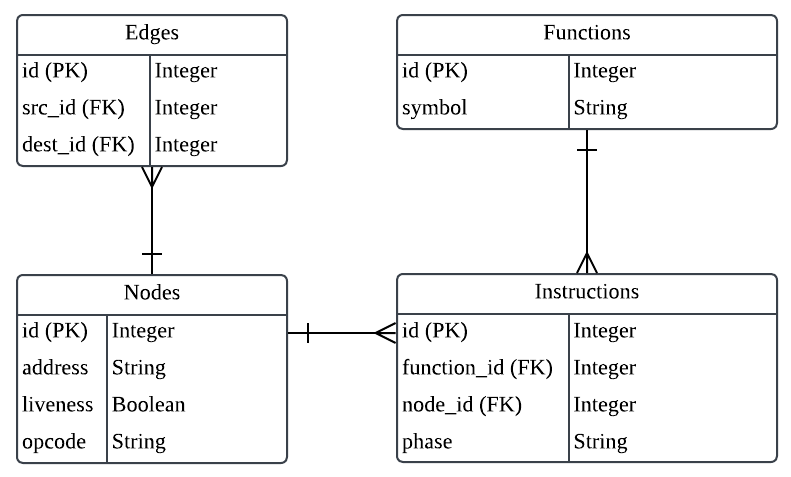}
\caption{Simplified Entity Relationship (ER) Diagram}
\label{fig:er_diagram}
\end{figure}
Figure~\ref{fig:er_diagram} presents a simplified version of the Entity-Relationship (ER) diagram for our database schema. The database consists of ten interrelated tables, each representing a specific component or transformation involved in the IR. However, for the purposes of visualization, we focus on the most essential tables. The \emph{Nodes} table stores metadata for each IR node, such as its unique ID, opcode, and alive status. The \emph{Edges} table represents source-to-destination relationships between nodes, enabling reconstruction of the IR graph structure. The \emph{Functions} table maps function IDs to human-readable function symbols. The schema's core is the \emph{Instructions} table, which logs all execution instructions that accessed or modified IR nodes. This table is key to identifying which nodes were created or transformed during specific optimization phases.


\subsection{Visualization Plan}
\begin{figure}[h]
\centering
\includegraphics[width=\columnwidth]{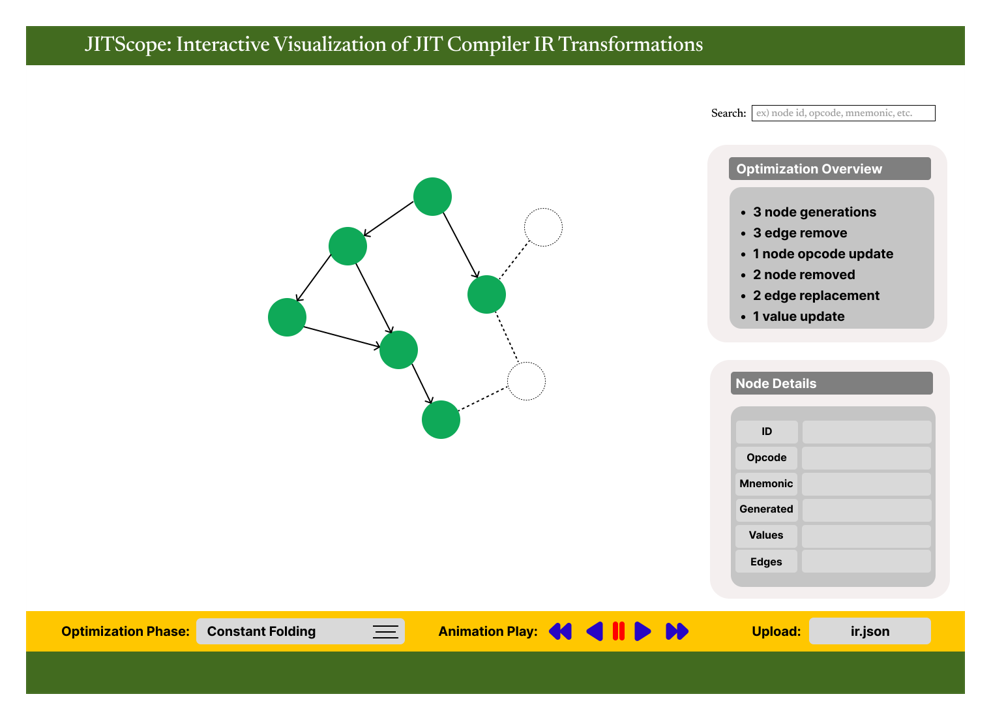}
\caption{Visualization Concept}
\label{fig:figma_ui}
\end{figure}


We are actively designing and implementing the front-end of JITScope
The current concept interface (Figure~\ref{fig:figma_ui}) illustrates our intended features: a central node-link graph showing the IR structure, a control panel at the bottom of the page, and the IR information display panel on the right.

\subsection{Center: IR Graph Visualization}
Figure~\ref{fig:figma_ui} shows the naïve visualization of the IR, i.e., nodes are laid out with no specific rules other than to avoid overlapping. The graph is to be interactive, i.e., nodes and edges (1) can be selected to view the details, (2) nodes are draggable, and (3) nodes and edges can be hidden or revealed by the user. Thus, our goal is to identify the most appropriate layout to show the relationship among the nodes and provide interactive features. At the same time, it is most intuitive for the developers to understand the visualized graph.

\subsection{Bottom: Control Panel}
At the bottom, the system provides a control option for the developers to interact with the IR.
\begin{enumerate}
    \item {\bf Dropdown Selector}: Using this selector, the users can switch around the phases to the changes made to the IR from one phase to the other in the central view of the IR graph.
    \item {\bf Animation keys}: Using the keys, the users can playback controls for animating transformations over time. The users should be able to play, pause, and fast forward/backward the animation.
    \item {\bf Upload}: A user can upload an IR JSON file. The users can only upload one file at a time.
\end{enumerate}

\subsection{Right: Information Panel}
On the right side of the interface, three key components support interaction and analysis:
\begin{enumerate}
    \item {\bf Search Bar}: This input field allows users to locate specific nodes within the currently selected optimization phase using attributes such as node ID, opcode, or mnemonic. The corresponding node is highlighted when a match is found, while the remaining nodes are visually de-emphasized (grayed out). This approach preserves the overall graph context, allowing users to manually explore other relevant nodes.
    \item {\bf Optimization Overview Panel}: This section summarizes the transformations applied to the IR during the current optimization phase. It includes counts of node generations, removals, opcode updates, value updates, and edge modifications (e.g., additions, removals, replacements), offering a quick, high-level snapshot of activity during that phase.
    \item {\bf Node Details Panel}: This panel displays detailed metadata for the selected node. It includes the node’s ID, opcode, mnemonic, a flag indicating whether it was generated during the current phase, and any associated value or edge changes.
\end{enumerate}

Although these are our initial plans to visualize, the details that should be displayed to aid the developers effectively in real development settings must be thoroughly studied. Our goal is to display only the most effective details for the developers while keeping the view minimal, so we can easily test the visualization and minimize the users' stress in using the system. To accomplish the goals, we plan to conduct surveys and interviews with the JIT compiler developers. The details of how this survey and interview would look are still in discussion.

\section{Brief Overview of Evaluation Plan}\label{sec:evaluation}
The evaluation of our visualization framework will focus on its usability and interpretability.
Our target user group is software developers working closely with JIT compilers and under-the-hood architecture. We propose three central research questions to guide this evaluation: 
\begin{enumerate}
    \item Can users accurately track how individual IR nodes evolve across different compiler phases?
    \item Is the resulting node-link graph readable and interpretable, especially for medium-sized IR datasets?
    \item Does the inclusion of phase-aware filtering enhance a user's ability to understand the nature and sequence of JIT compiler transformations?
\end{enumerate}

To address these questions, we will employ a combination of qualitative and quantitative metrics. Node clarity will be assessed based on the readability of tooltips, the legibility of node labels, and the visual coherence of node placement. User correctness will be measured through task-based evaluation, where users are asked to complete specific challenges such as identifying the phase during which a node’s value changed or determining which function last accessed a given node. These tasks will allow us to quantify the accuracy and efficiency with which users can extract meaningful insights from the visualization. 


\section{Discussion}\label{sec:discussion}
%

\subsection{IR JSON}
A key challenge in building our visualization was handling the complexity of phase mapping within the input IR JSON file. It required nuanced traversal logic over function IDs to correctly associate each instruction with its corresponding optimization phase. The IR data is generated using the tool developed by Lim \emph{et al.}~\cite{lim2022modeling}. Their data model was designed for bug localization, and as such, it preserves deeply nested structures and exhaustive optimization metadata—not all of which are necessary for visualization. Our initial effort focused on identifying which portions of this rich dataset are most relevant for visual exploration and designing an efficient extraction pipeline that retains semantic clarity while reducing structural complexity.

\subsection{Front-End Visualization}
D3’s layout constraints (e.g., non-hierarchical data handling) limited our ability to use ideal formats like edge bundling. We also faced limited time to verify the correctness of visualization outputs. Some visual clusters might represent inactive nodes, and hover behavior needs refinement for crowded graphs. However, the modularity of our system will support future improvements. Meanwhile, the current prototype has not yet been adjusted fully for scalability, and further considerations may be added upon testing more dense IR networks.

\section{Related Work}\label{sec:related-work}
%

The closest work in visualizing JIT compiler IRs is the work by Lim and Kobourov~\cite{lim2021jitgraphs}. Their approach visualizes the entire IR structure in a single, cohesive layout using the metro map metaphor~\cite{jacobsen2021metrosets}, providing a high-level overview of the IR graph. However, the visualization is static and does not capture the evolution of the IR across different compiler phases. As a result, it offers limited support for analyzing phase-specific transformations, which are often crucial for debugging and understanding JIT optimization behavior.

Reiss introduced Jive, a real-time visualization system for Java programs~\cite{reiss2003visualizing}. The system visualizes classes used during program execution by representing them as boxes, allowing the users to observe runtime behavior. While this work shares a similar goal of visualizing execution-related components, its focus is on illustrating the structure and behavior of the input program itself. It does not address the internal behavior of the Java runtime environment, such as the interpreter or the JIT compiler, which are the primary focus of our work.

Several visualization tools are designed to support debugging~\cite{gnuDDD,baskerville1985graphic}. These tools commonly use interactive box-and-arrow diagrams to represent the structure and behavior of the input program visually. Many offer advanced features such as pausing execution, stepping through program states, and clicking on visual elements to generate additional views—such as performance charts or data flow summaries. While these tools share the goal of improving the developer's insight through visual interaction, their focus is primarily on visualizing the input program itself rather than visualizing the internal behavior of compilers.

\section{Conclusion}\label{sec:conclusion}
JITScope presents a modular, interpretable framework for visualizing JIT compiler IR evolution. From JSON to SQL to interactive IR graphs, our pipeline enables the exploration of node behaviors, value changes, and phase transitions. Our front-end is still in the active implementation stage. We aim to refine our hierarchical graph design in future iterations, test usability with compiler engineers, and integrate backend–frontend live interaction.

\section*{Acknowledgements}\label{sec:ack}

This work was supported in part by Davidson College.

\bibliographystyle{IEEEtran}
\bibliography{references}



\end{document}